\documentclass[10pt]{iopart}
\pdfoutput=1
\usepackage{iopams}
\usepackage{epsfig,cite}
\newcommand{\be}{\begin{equation}}
\newcommand{\ee}{\end{equation}}
\newcommand{\bd}{\begin{displaymath}}
\newcommand{\ed}{\end{displaymath}}
\newcommand{\BE}{\begin{eqnarray}}
\newcommand{\EE}{\end{eqnarray}}

\newcommand{\id}{{\rm 1\!\! I}}

\newcommand{\bn}{\ensuremath{\mathbf{n}}}

\newcommand{\bx}{\ensuremath{\mathbf{x}}}

\newcommand{\avg}[1]{\left\langle{#1}\right\rangle}

\newcommand{\boldeta}{{\mbox{\boldmath $\eta$}}}
\newcommand{\boldvarphi}{{\mbox{\boldmath $\varphi$}}}

\begin{document}

\title{Fluctuations in meta-population exclusion processes}

\author{Tobias Galla}

\address{
Theoretical Physics, School of Physics and Astronomy, The University of Manchester, Manchester M13 9PL, United Kingdom
}

\begin{abstract}
We introduce a meta-population version of models of asymmetric exclusion models, consisting of a spatial arrangement of patches. Patches are of a specific size, indicating the maximal number of particles they can hold. We use an expansion in the inverse patch size to calculate the spectral properties of fluctuations in such systems. This provides a systematic derivation from first principles of effective Langevin descriptions discussed in the literature. We apply our approach to the totally asymmetric simple exclusion process, to variants with an overall constraint on the total particle number and to a two-species exclusion model. The theory provides semi-analytical results, these are confirmed in numerical simulations, and give good approximations to conventional exclusion models. These are recovered when the patch size is set to unity.
\end{abstract}


\ead{\tt Tobias.Galla@manchester.ac.uk}


\section{Introduction}
The asymmetric exclusion process (ASEP) is one of the most studied models in non-equilibrium statistical physics. Originally introduced in the context of molecular transport \cite{macdo1,macdo2, shaw,dong,frey}, it has found wide applications not only to model biological systems, but also to the modelling of pedestrian motion and the formation of traffic jams \cite{nagel,schad, schad2,wolki,helbing}. Most ASEP models describe a chain of cells, each of which can either be vacant or filled by one particle. Particles are injected stochastically at one end of the chain and then propagate from one cell to the next according to a stochastic rule. Crucially they can only move ahead if the subsequent cell is not occupied, otherwise their motion is blocked until the cell ahead becomes vacant. Stochastic ejection occurs at the other end of the chain. A comprehensive review of the theory and applications of exclusion processes and related models can be found in \cite{schadschneider}.  In statistical physics exclusion processes represent a widely studied class of driven lattice gases\cite{derrida,schutz,derrida2,derrida3,derridaevans,evans,blythe}, a variety of different methods have been used to characterize their phase behaviour, and to derive exact or approximate solutions \cite{derrida,schutz, derridaevans,schadschneider,schad2,blythe,essler1,essler2,essler3,kolo}.

\vspace{0.5em}

Formulating a mean-field theory of exclusion processes is relatively straightforward, see \cite{macdo2,schadschneider}. At the same time it can give accurate insights into the basic phenomena displayed by such models. The mathematics required to carry out exact analyses on the other hand is intricate, so that it is desirable to develop approaches which systematically improve on mean field descriptions, but do not require an overly involved mathematical apparatus. Fluctuation effects in ASEP models have for example been studied by an effective Langevin description in \cite{frey2,zia0,zia1,zia2,cookdong}.  While successful in describing the spectral properties of fluctuations about the mean field theory the precise form of these equations is often not derived from first principles. Instead they are formulated phenomenologically, frequently some of the resulting coefficients are effectively treated as fitting parameters.

The purpose of the present work is to discuss a more systematic approach to deriving these effective Langevin equations. We start from what we will refer to as a `meta-population' version of the totally asymmetric exclusion process. The term `meta-population' goes back to Levins \cite{levins} who used it to describe `populations of populations'. We use it in the context of the ASEP to describe models in which each cell can be occupied by more than one particle. Each cell then becomes a `patch' in which a population of particles can reside. Particles may hop from one patch to another according to rules to be specified below. This introduces an interaction between the patches, and the aggregate system constitutes a population of interacting populations.

More specifically, we consider a model in which each site can contain up to $\Omega$ particles, where $\Omega$ is a fixed positive integer. Instead of being simply occupied or vacant, as in the conventional totally asymmetric simple exclusion process (TASEP), sites in this extended model are characterized by a filling factor $x_i$, which can take values $x_i=0,1/\Omega,2/\Omega,\dots,1$. Hopping from patch $i$ to $i+1$ then occurs with a rate proportional to $x_i(1-x_{i+1})$, in particular no hopping is possible if the cell ahead is fully occupied. Our semi-analytica approach is then based on a systematic expansion of the master equation description in powers of $\Omega^{-1/2}$.

While introducing a model with multiple occupancy might seem to constitute a significant departure from the original TASEP we believe there is considerable merit in studying the meta-population model within a van Kampen expansion picture.  This is motivated by the following results: (i) In the limit $\Omega=1$ the model reduces to the standard TASEP; (ii) As we will show, to lowest order in the expansion in $\Omega^{-1/2}$ (i.e. taking the limit $\Omega\to\infty$) the model reproduces the mean field equations usually written down for the TASEP, thus the model interpolates between the conventional TASEP and its mean field theory; (iii) Taking the expansion to next-to-leading order reproduces the Langevin dynamics proposed and studied in \cite{zia0,zia2,cookdong}; (iv) Crucially our approach is fully controlled, and allows one to state the limitations of the description in terms of Gaussian random processes; (v) The relevant coefficients in the Langevin (or Ornstein-Uhlenbeck) dynamics are derived from first principles, they do not need to be obtained by a fit from simulation data, our analysis may hence also serve as a starting point for a better understanding of the `serious renormalization' of diffusion constants and noise strengths reported in \cite{zia0}.  We apply these methods first to the basic TASEP and then to what is referred to as the constrained TASEP \cite{zia1,zia2}. As a further application we study a two-species exclusion model, to our knowledge no calculations based on Langevin approaches have been reported to date for this variant.

\vspace{0.5em}

The remainder of this paper is organised as follows: In Sec. \ref{sec:model} we introduce the meta-population variant of the single-species TASEP model. A systematic expansion in the inverse capacity of the individual cells is the carried out in Sec. \ref{sec:vk}. Results are compared against simulations in Sec.  \ref{sec:sim}, both for the standard totally asymmetric simple exclusion process, and for an exclusion process with an overall constraint on the total particle number, see also \cite{zia1,zia2}. In Sec. \ref{sec:multi} we then address a specific two-species exclusion process, before we draw our conclusions and give an outlook on potential future research in Sec. \ref{sec:concl}.
\section{Model and master equation description}\label{sec:model}

\begin{figure*}[t]
\vspace{2em}
\centering
\vspace{-3em}
\includegraphics[scale=0.35]{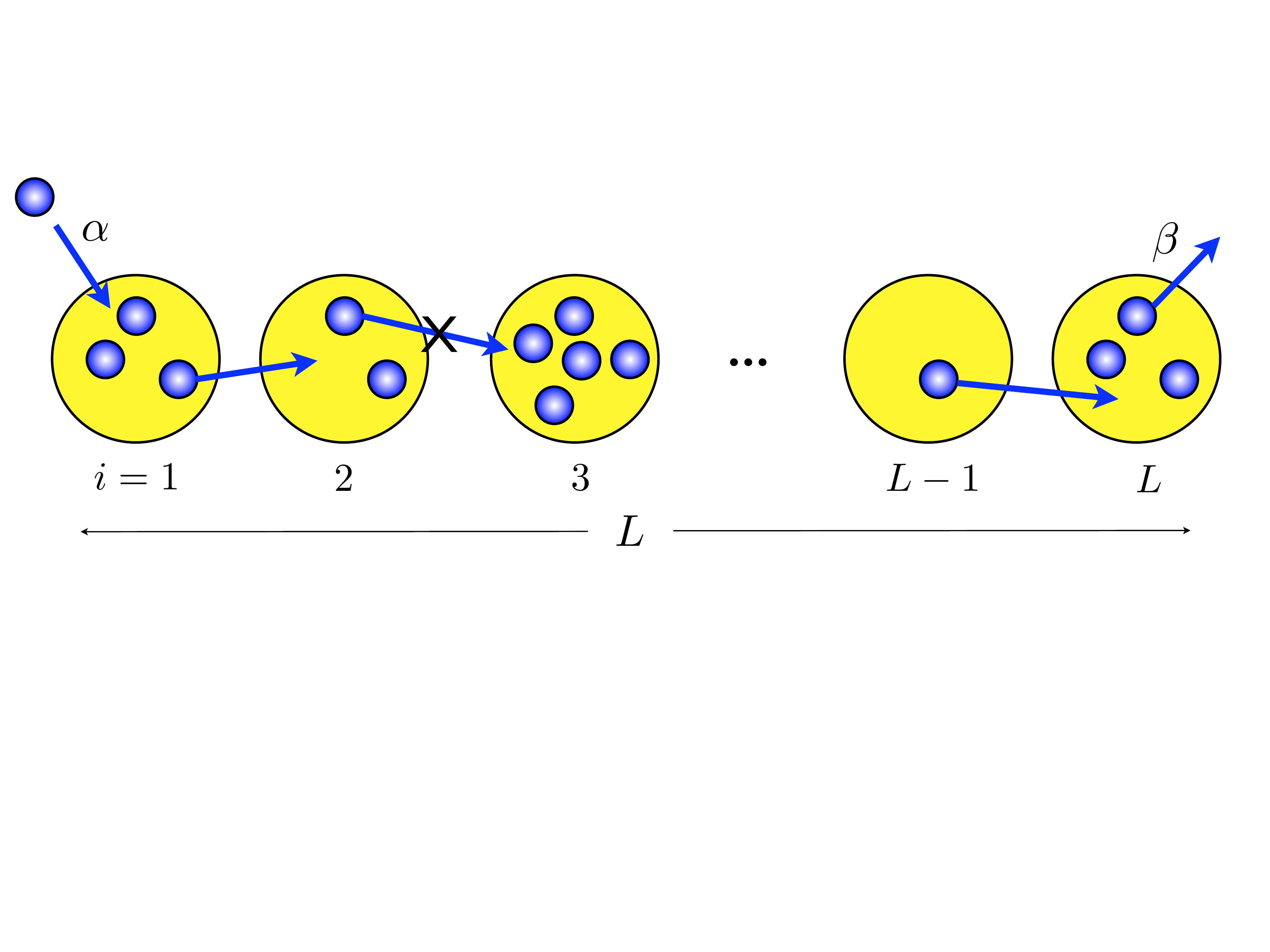} 
\vspace{-7em}
\caption{An illustration the `meta-population' asymmetric exclusion process with a capacity of up to $\Omega=5$ particles per cell. The arrows depict possible transitions for the given configuration, hopping from cell $2$ to cell $3$ is not allowed, as cell $3$ is fully occupied.}
\label{fig:fig1}
\end{figure*}

Our model system consists of $L$ patches (or urns), labelled $i=1,\dots,L$. Each urn can accommodate up to $\Omega$ particles, as shown in Fig. \ref{fig:fig1}. We will write $n_i(t)\in\{0,\dots,\Omega\}$ for the number of particles in urn $i$ at time $t$. The state of the system at any given time is fully characterized by the occupation numbers $\bn=(n_1,n_2,\dots,n_L)\in\{0,\dots,\Omega\}^L$. 

The continuous-time dynamics we will consider is defined by the following transition rates (from state $\bn$ to state $\bn'$):
\BE\label{eq:t}
T_0(\bn'|\bn)&=&\alpha(n)\frac{\Omega-n_1}{\Omega}\delta_{{n'_1},{n_1}+1},\nonumber \\
T_i(\bn'|\bn)&=&\frac{n_i}{\Omega}\frac{\Omega-n_i}{\Omega}\delta_{{n'_i},{n_i}-1}\delta_{{n'_{i+1}},{n_{i+1}}+1}, ~~i=1,\dots,L-1,\nonumber \\
T_L(\bn'|\bn)&=&\beta\frac{n_L}{\Omega}\delta_{{n'_L},{n_L}-1},
\EE
where $\delta_{n',n}$ is the Kronecker delta, i.e. $\delta_{n',n}=1$ for $n=n'$ and $\delta_{n',n}=0$ otherwise. The first reaction rate, $T_0$, here describes injections of particles into urn $i=1$. Such injection is only possible if this urn is not fully filled already (i.e. only if $n_1<\Omega$), and it occurs with a rate proportional to the model parameter $
\alpha\geq 0$. In order to keep the setup sufficiently general to allow for ASEP models with an overall constraint on the particle number in the system we assume that $\alpha=\alpha(n)$, where $n=\sum_i n_i$. For example we will consider models in which $\alpha(n)$ tends to zero when $n$ approaches a maximal capacity. The transition rates  $T_i$, $i=1,\dots,L-1$ correspond to moving a particle from urn $i$ to $i+1$, again this is only possible if the destination urn is not fully occupied already. The quantity $T_{L}$ finally is the rate with which particles are ejected from the last urn, $i=L$, and is proportional to the second main model parameter $\beta\geq0$.

The time evolution of the probability $P_t(\bn)$ of finding the system in state $\bn$ at time $t$ is then given by the following master equation
\be\label{eq:master}
\partial_t P_t(\bn)=\sum_{i=0}^L\sum_{\bn'\neq\bn} \bigg[T_i(\bn|\bn')P_t(\bn')-T_i(\bn'|\bn)P_t(\bn)\bigg].
\ee
Simulations are carried out using the celebrated Gillespie algorithm \cite{gillespie,gillespie2}, which allows one to generate realizations of the stochastic process described by Eq. (\ref{eq:master}).

\section{System-size expansion and calculation of power spectra}\label{sec:vk}
\subsection{Deterministic limit}
Deterministic mean field equations can be obtained by multiplying the above master equation by $\bn$ on both sides and then subsequently summing over $\bn$. Writing $x_i(t)=\Omega^{-1}\avg{n_i(t)}=\Omega^{-1}\sum_{\bn} n_i P_t(\bn)$ one finds
\be
\dot x_i=T_{i-1}^\infty(\bx)-T_i^\infty(\bx), ~~ $i=1,\dots,L$,
\ee
where we have written $T_0^\infty(\bx)=\alpha^\infty(x)(1-x_1)$, as well as $T_i^\infty(\bx)=x_i(1-x_{i+1})$ for $i=1,\dots,L-1$. The term describing ejection from cell $L$ is given by $T_L^\infty(\bx)=\beta x_L$. We have here introduced $\alpha^\infty(x)$ for the injection rate in the deterministic limit, where $x=\Omega^{-1}\sum_{i=1}^L \avg{x_i}$. It is important to note that this is a heuristic derivation and that an approximation has been made to factorize quantities such as $\avg{x_ix_{i+1}}$ into $\avg{x_i}\avg{x_{i+1}}$. The resulting mean-field dynamics can then be written as
\BE
\dot x_1&=&F_1(\bx):=\alpha(x) (1-x_1)-x_1(1-x_2), \nonumber \\
\dot x_i&=&F_i(\bx):= x_{i-1}(1-x_i)-x_i(1-x_{i+1}), ~~~ i=2,\dots,L-1 \nonumber \\
\dot x_L&=&F_L(\bx):= x_{L-1}(1-x_L)-\beta x_L.\label{eq:mf}
\EE
Some explanation regarding the nature of these equations is here appropriate. While these equations are easy to write down and intuitive in their interpretation it is a-priori unclear what exactly the underlying approximations are that have been made to arrive at them. The above mentioned factorization for example is a consequence of systematically neglecting all fluctuations and e.g. replacing the distribution of realized values of $n_i$ by a delta function at its mean value, i.e. $P_t(\bn)\rightarrow \delta(\bn-\avg{\bn(t)})$ The expansion technique in powers of $\Omega^{-1/2}$ we will discuss below can be useful to understand these issues. The above deterministic equations are precisely the outcome of the lowest order of the expansion, they are therefore exact in the limit $\Omega\to\infty$. 

In this paper we are mostly interested in the stochastic dynamics in the stationary regime, and as we will see the properties of fluctuations in this regime can be computed from the long-time behaviour of the above deterministic equations. Numerically integrating these equations, either for constant $\alpha$, or for $\alpha=\alpha(x)$ in the case of constrained the TASEP, we generally find that these equations approach a fixed point $\bx^*=(x_1^*,\dots,x_L^*)$ asymptotically. These fixed point values are evaluated numerically, and then used for the analysis of stochastic effects, as described below.
\subsection{Leading-order corrections}
Expanding the above master equation in powers of $\Omega^{-1/2}$ allows one to systematically characterize fluctuation effects about the above mean-field dynamics. This technique is known as the system-size expansion and goes back to van Kampen \cite{kampen}\footnote{We here stress that the expansion parameter in our analysis is the inverse square root of the capacity $\Omega$ of each cell. The resulting theory thus applies in the limit of large, but finite $\Omega$. The number of cells, $L$, in the system is a separate model parameter, and remains finite throughout.}. The starting point is the decomposition
\be\label{eq:dec}
\frac{n_i}{\Omega}=x_i(t)+\frac{\xi_i(t)}{\sqrt{\Omega}},
\ee
of the state $\bn$ of the system into a deterministic part, $\bx(t)$ and fluctuations $\bxi(t)$. This procedure is straightforward and has been applied to spatial and non-spatial systems for example in \cite{kampen,alan,alonso,kuske,reichenbach, mobilia,nunes,pineda,gallaprl,tommaso}, the technical details of such calculations are extensively described in \cite{kampen} so that we do not report the intermediate steps here. In fact the result of the expansion can be written down directly, using the general results decribed in the Appendix of \cite{bladon}. 

To leading order in the expansion one obtains the deterministic equations described above (Eqs. (\ref{eq:mf})). To first order in $\Omega^{-1/2}$ one finds a linear Fokker-Planck equation of the form
\be\label{eq:fp}
\partial_t \Pi(\bxi)=-\sum_i \partial_i \left[\sum_j J_{ij}\xi_j\Pi(\bxi)\right]+\frac{1}{2}\sum_{ij}\partial_i \partial_j \left[B_{ij}\Pi(\bxi)\right],
\ee

 describing the evolution of fluctuations $\bxi$ about the deterministic trajectory. We have here written $\partial_i\equiv \partial/\partial \xi_i$. The matrices $\mathbb{J}$ and $\mathbb{B}$ are given by
 \be\label{eq:j}
 J_{ij}=\frac{\partial F_i}{\partial x_j}
 \ee
 and
 \be\label{eq:b}
 B_{ij}=-T_{i-1}^\infty\delta_{i-1,j}+[T_i^\infty+T_{i-1}^\infty]\delta_{ij}-T_i^\infty\delta_{i+1,j}, ~~ i,j=1,\dots,L,
 \ee
and where all other elements of $\mathbb{B}$ vanish. We are only interested in the behaviour at large times, so all expressions in Eqs. (\ref{eq:j}) and (\ref{eq:b}) are to be evaluated at the deterministic fixed point. The Fokker-Planck equation (\ref{eq:fp}) is equivalent to the following set of Langevin equations
\be
\dot \xi_i(t)=\sum_j J_{ij}\xi_j(t)+\eta_i(t),
\ee
where $\boldeta(t)$ is Gaussian white noise of mean zero and with correlations
\be
\avg{\eta_i(t)\eta_j(t')}=B_{ij}\delta(t-t')
\ee
among its components.
 This can be written as
\BE
\dot \xi_1(t)&=&-\alpha[x^*(t)]\xi_1(t)-[1-x^*_2(t)]\xi_1(t)+x^*_1(t)\xi_2(t)\nonumber \\
&&+[1-x^*_1(t)]\alpha'[x^*]\sum_i\xi_i(t)+\eta_1(t) \nonumber \\
\dot \xi_i(t)&=&[1-x^*_{i}(t)]\xi_{i-1}(t)-x^*_{i-1}(t)\xi_i(t)\nonumber\\
&& -[1-x^*_{i+1}(t)]\xi_i(t)+x^*_i(t)\xi_{i+1}(t)+\eta_i(t), ~~~i=2,\dots,L-1  \nonumber \\
\dot \xi_L(t)&=&(1-x^*_L(t))\xi_{L-1}(t)-x^*_{L-1}\xi_L(t)-\beta\xi_L(t)+\eta_L(t),
\EE
(with $\alpha'(x)=d\alpha/dx$), or in more compact form as
\be
\dot {\bxi}=\mathbb{J}\bxi+\boldeta,
\ee
leading to
\be
\left[i\omega\id-\mathbb{J}\right]\widetilde{\bxi}(\omega)=\widetilde{\boldeta}(\omega)
\ee
in Fourier space.
This can be inverted straightfowardly, and one obtains the power spectrum
\be\label{eq:specij}
\avg{\widetilde \xi_i(\omega)\widetilde \xi_j(\omega')}=\left[(i\omega\id-\mathbb{J})^{-1}B(-i\omega\id-\mathbb{J}^T)^{-1}\right]_{ij}\delta(\omega+\omega').
\ee
 This set of expressions in principle contains full information about the temporal auto-correlations and cross-correlations of the components of $\bxi$, and hence describes the properties of fluctuations about the deterministic model to the full (in the Gaussian approximation we have made truncating the van Kampen expansion after the sub-leading term). We will here mostly focus on the fluctuations of the total number of particles in the system about the deterministic value $x=\sum_i x_i^*$. These are given by $\xi(t)=\sum_i \xi_i(t)$ and their power spectrum is obtained as $\avg{\widetilde \xi(\omega)\widetilde \xi(\omega')}=\sum_{ij}\avg{\widetilde \xi_i(\omega)\widetilde \xi_j(\omega')}$. One finds
\be
\avg{\widetilde \xi(\omega)\widetilde \xi(\omega')}=P(\omega)\delta(\omega+\omega'),
\ee
where
\be\label{eq:specx}
P(\omega)=\sum_{ij} \left[(i\omega\id-\mathbb{J})^{-1}B(-i\omega\id-\mathbb{J}^T)^{-1}\right]_{ij}.
\ee
We stress that these are the spectra of $\xi(t)$, recalling Eq. (\ref{eq:dec}) it might be more appropriate to refer to them as re-scaled power spectra, given that a factor of $\Omega$ has been already scaled out in the spectra. As a result the expression in Eq. (\ref{eq:specx}) does not depend on $\Omega$. In our further analysis this expression is evaluated numerically\footnote{To this end we have carried out the $L\times L$ matrix inversions and multiplications numerically while varying $\omega$. This was done mostly for $L=51$. At larger chain lengths it might be advisable to diagonalize $\mathbb{J}$ once and then to work in the corresponding eigenspace. It is then not required to invert $L\times L$ matrices for all values of $\omega$ tested, instead one single matrix inversion is then sufficient.} and will be compared against simulations for a several different variants of the asymmetric exclusion process.
\section{Test against simulations}\label{sec:sim}
In this section we will compare results obtained from our analytical calculations against numerical simulations of the exclusion process. We first address the standard TASEP and then the constrained TASEP.
\subsection{Standard TASEP}
\begin{figure*}[t!!]
\vspace{2em}
\centering
\vspace{-1em}
\includegraphics[scale=0.55]{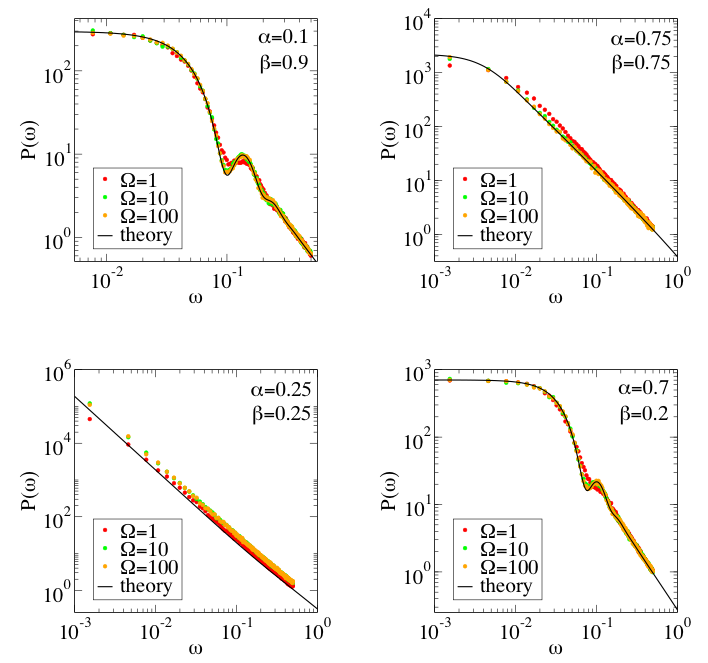} 
\vspace{0em}
\caption{(Colour online) Power spectrum of fluctuations of the total number of particles in the system about their deterministic value. Each of the four panels corresponds to different combinations of $\alpha$ and $\beta$, corresponding to the LD phase (upper left panel), the MC phase (upper right), the coexistence phase (lower left) and the HD phase (lower right). In each panel we show the theoretical prediction, obtained from Eqs. (\ref{eq:specij},\ref{eq:specx}), and results from simulations at different cell volumes $\Omega$, averaged over at least $10$ independent runs. The number of cells in the system is $L=51$.}
\label{fig:fig2}
\end{figure*}
In the standard TASEP there is no constraint on the overall number of particles in the system, injection only depends on the occupancy of the first cell, the injection rate assumes a constant value $\alpha\neq\alpha(n)$. The dynamics is hence specified by the two parameters $\alpha$ and $\beta$, the latter being the constant ejection rate at the end of the chain. This system is known to exhibit several dynamic phases, see e.g. \cite{schadschneider} and references therein: (i) the so-called high-density phase (HD) at $\alpha>\beta$ and $\beta<1/2$, (ii) the low-density phase (LD) at $\beta>\alpha$ and $\alpha<1/2$, (iii) a co-existence phase at $0<\alpha=\beta<1/2$ and (iv) the so-called maximum current phase (MC) at $\alpha,\beta>1/2$.

We show results for the power spectrum of the fluctuations of the total particle number in the four different phases in Fig. \ref{fig:fig2}. In the co-existence and MC phases we find algebraic decay of the power spectra, as already observed in \cite{zia0,zia2}. In the LD and HD phases the power spectra again display power-law decay, but modulated by damped oscillations, compare again with \cite{zia0,zia2}. 

In all four phases we find good general agreement between the analytical predictions based on van Kampen's system-size expansion and numerical simulations.  Naturally, the theory compares better against simulations when $\Omega$ is large (see the results for $\Omega=100$ in Fig. \ref{fig:fig2}), one should here keep in mind that the system-size expansion approach is valid in the limit of large, but finite $\Omega$. Small deviations can never fully be eliminated, see e.g. Fig. \ref{fig:fig3} at small frequencies. We attribute these to either fluctuations, limitations in taking the $\Omega\to\infty$ limit (simulations are necessarily carried out at finite $\Omega$) and to other numerical effects relating to taking the Fourier transform of finite time series. As seen in Fig. \ref{fig:fig2} the agreement between theory and simulations is reasonably good also for $\Omega=1$, corresponding to the original TASEP, in which each cell can hold at most one particle. We re-iterate again that no fitting procedure has been carried out at any step of our analysis. Occasionally one finds non-mononotic behaviour of the re-scaled spectra as a function of $\Omega$, see e.g. the lower left panel of Fig. \ref{fig:fig2}. This is presumably due to effects not captured by the expansion to first order in $\Omega^{-1/2}$ (which is inherently based on the assumption that fluctuations are of order $\Omega^{-1/2}$, and predicts re-scaled spectra hence do not depend on $\Omega$). We speculate that going to higher orders in the expansion might be able to describe these non-monotonicities. We note though that the agreement between theory and simulations is less good in the co-existence phase (lower left panel in Fig. \ref{fig:fig2}, the deviation can actually amount to up to a factor of two at certain frequencies). This is either due to the fact that we have not fully reached the large-$\Omega$ limit, or due to the intrinsic dynamics of this phase. Mean field solutions typically display a `kink' in particle density (see e.g. \cite{derrida}), and we find that the location of this domain wall can depend on initial conditions used to find the fixed point of the deterministic equations, and as stated in \cite{derrida} on the precise nature in which the limit $\alpha\to\beta$ limit is taken\footnote{Surprisingly this does not seem to affect the resulting spectrum, which appears to be independent of the position of the domain wall.}. We cannot exclude that certain phenomena are at work here which we do not fully understand, and we would like to limit our conclusions for the co-existence phase to the observation that the theory qualitatively reproduces the shape of the power spectrum correctly. 
\subsection{Constrained TASEP}
In the constrained TASEP the effective injection rate $\alpha$ depends on the number of particles present in the system, and particles enter and leave at fixed rates. Other approaches have been considered e.g. in \cite{wood} where a gating process was introduced. We here follow \cite{zia2} who proposed a model with an overall constraint on the number of particles in the system. Specifically we will use
\be
\alpha(n)=\alpha_0\tanh\left[\frac{n_m-n}{n_c}\right],
\ee
where $n$ is the total number of particles in the system, $n_m$ the maximum number of particles allowed in the system. The quantity $n_c$ is a cross-over parameter, defining the precise shape of the sigmoidal function $\alpha(n)$. Adapting this to the meta-population exclusion process, in which each cell can hold up to $\Omega$ particles, the effective injection rate is given by
\be
\alpha(n)=\alpha_0\tanh\left[\frac{\rho_m-\frac{n}{\Omega L}}{\rho_c}\right],
\ee
where $\rho_m$ is the maximally allowed density of particles, and where $\rho_c$ is an appropriately normalized equivalent of $n_c$.

\begin{figure}[t]
\vspace{2em}
\centering
\vspace{-1em}
\includegraphics[scale=0.45]{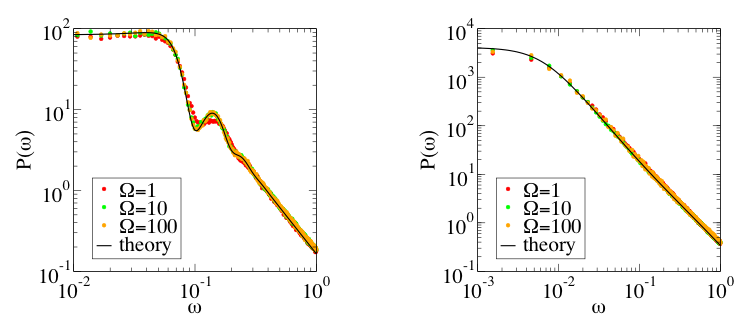} 
\vspace{0em}
\caption{(Colour online) Power spectra of fluctuations of the total particle density in the constrained TASEP. {\bf Left:} $\beta=0.9$, $\alpha_0=0.3, \rho_{m}=0.205, \rho_c=0.3$. {\bf Right:}  $\beta=0.3$, $\alpha_0=0.7, \rho_{m}=0.8, \rho_c=0.7$. In each panel we show the theoretical prediction, obtained from Eqs. (\ref{eq:specij},\ref{eq:specx}), and results from simulations at different cell volumes $\Omega$, averaged over at least $10$ independent runs. The number of cells in the system is $L=51$. \label{fig:fig3}}
\end{figure}

Results are shown in Fig. \ref{fig:fig3} for two sets of the model parameters, chosen to correspond to the values used in \cite{zia2}. As seen in the figure two distinct regimes can again be identified, one with an algebraic decay of the power spectrum and another with additional oscillatory modulations. The agreement between theory and simulations is again good in, with only relatively small deviations at small values of $\Omega$. It is here appropriate to comment briefly on one difference between our approach and that of \cite{zia2}. The authors of \cite{zia2} phenomenologically derive Langevin equations not too dissimilar from the ones we here obtain using a more systematic approach based on the cell-size expansion. A further difference between our work and that of \cite{zia2} concerns the details of the deterministic fixed point used to compute the power spectra of calculations. In \cite{zia2} conditions are chosen such that a homogeneous density profile can be assumed, i.e. $x_i^*=x_j^*$ for all $i,j$. If this assumption is made the system effectively becomes translation invariant, and can be diagonalized in Fourier space (the Fourier transform would here be carried with respect to position space, i.e. cell numbers). In our calculation we do not make this assumption, but consider a general deterministic fixed point $\bx^*=(x_1,\dots,x_{L}^*)$, and as a consequence the system of $L$ resulting Langevin equations can not easily be simplified. Computing the power spectra as detailed in Eqs. (\ref{eq:specij},\ref{eq:specx}) therefore requires more computational resources than for the approach taken in \cite{zia2}. Due to this, and to the fact that we simulate models of up to $\Omega=100$ particles per site, results in our figures are mostly limited to $L=51$, whereas much larger systems are considered in \cite{zia2}. The different number of cells in the system explain the differences between the figures in \cite{zia2} and ours. This applies to the standard TASEP as well as to the model with a constrained total number of particles.
\section{Two-species TASEP}\label{sec:multi}
\subsection{Definition}
As a third example we consider a TASEP with two distinct species of particles \cite{derrida3}. For $\Omega=1$ each cell can then either be empty $(0)$, occupied by a particle of type $1$, or occupied by a particle of type $2$. Particles of type $1$ behave the same way as particles in the conventional TASEP do, they hop ahead to the subsequent cell if that cell is not occupied by a particle of type $1$. In other words, particles of type $1$ take no notice of the presence or absence of type-$2$ particles. One has the processes $10\longrightarrow 01$ and $12\longrightarrow 21$. Particles of type $2$ on the other hand are barred from moving if the cell ahead is occupied by a particle of any type, they only move if the cell ahead is vacant. Particles of type $2$ thus experience an interaction with particles of type $1$, one has $20\longrightarrow02$, but no movement in a configuration of the form $21$. As one key difference to the two models considered before we here choose periodic boundary conditions. There is no injection or ejection of particles, the total number of particles of each type is conserved throughout. This setup makes the system translation invariant, hence simplifying the calculation of the spectral properties of fluctuations. This will be detailed below.
 
\begin{figure}[t]
\vspace{2em}
\centering
\vspace{-1em}
\includegraphics[scale=0.45]{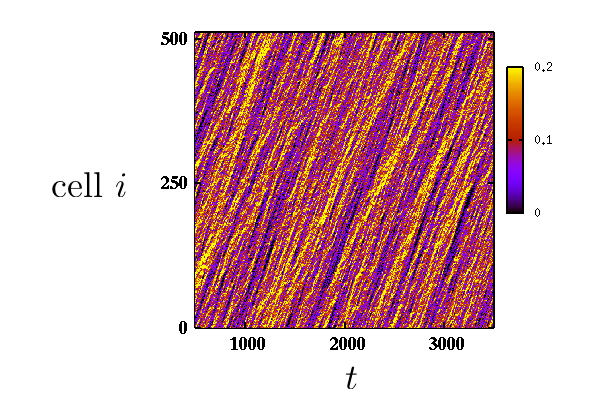} 
\vspace{0em}
\caption{(Colour online) Space-time diagram of the density of particles of type $2$ obtained from a single simulation run of the two-species TASEP. The system consists of $512$ cells, model parameters are $\rho_1=0.1, \rho_2=0.2$, each cell can hold up to $\Omega=10$ particles. The colour map indicates the density of particles of type $2$, $m_i(t)/\Omega$ in cell $i$ at time $t$.\label{fig:fig4}}
\end{figure}

This model is easily generalized to the `meta-population' case in which each cell can hold up to $\Omega$ particles total. If we denote the number of type-$1$ particles in cell $i$ by $n_i$, and the number of type-$2$ particles in that cell by $m_i$, then we have $n_i+m_i\leq \Omega$ at all times. The meta-population model is the defined by the transition rates
\BE
T_{i,1}=\frac{n_i(\Omega-n_{i+1}-m_{i+1})}{\Omega}, \nonumber \\
T_{i,2}=\frac{n_im_{i+1}}{\Omega}, \nonumber \\
T_{i,3}=\frac{m_i(\Omega-n_{i+1}-m_{i+1})}{\Omega},
\EE
corresponding to the processes $10\longrightarrow 01$, $12\longrightarrow 21$ and $20\longrightarrow 02$ respectively\footnote{In \cite{derrida3} these different processes are assumed to occur with independent and potentially different rates, for simplicity we here focus on the case of equal rate constants. Generalization is straightforward.}. Execution of a reaction with rate $T_{i,2}$ will for example lead to the update $n_i\to n_i-1, n_{i+1}\to n_{i+1}+1, m_i\to m_i+1, m_{i+1}\to m_{i+1}-1$. In order to capture the periodic boundary conditions we wish to address, expressions of the type $i+1$ and $i-1$ are here to be read `modulo $L$', where $L$ is the total number of sites in the ring. A simulation of this process is shown in Fig. \ref{fig:fig4}.
\subsection{Deterministic and stochastic analysis}
Denoting the concentration of particles of type $1$ in cell $i$ by $x_i=\frac{n_i}{\Omega}$, and that of particles of type $2$ by $y_i=\frac{m_i}{\Omega}$ on obtains the following deterministic dynamics in the limit $\Omega\to\infty$:
\BE
\dot x_i &=& -x_i(1-x_{i+1})+(1-x_i)x_{i-1}, \nonumber \\
\dot y_i &=& x_i y_{i+1}-x_{i-1}y_i+y_{i-1}(1-x_i-y_i)-y_i(1-x_{i+1}-y_{i+1}).\label{eq:2spmf}
\EE
Clearly, $x_i^*=\rho_1$, $y_i^*=\rho_2$ for all $i$ is a fixed point for all $\rho_1,\rho_2$. This reflects the fact that particle numbers are conserved and that the system is translation invariant. The densities $\rho_1$ and $\rho_2$ are indeed model parameters of the two-species TASEP, for obvious reasons we restrict their choice to non-negative values with $\rho_1+\rho_2\leq 1$.

It is again straightforward to carry out an expansion in the inverse cell size, calculations of this type for spatial systems can for example be found in \cite{lugo,butler1,butler2,tommaso2,tommaso}. To first order in $\Omega^{-1/2}$ one again obtains a set of Langevin equations describing the fluctuations about the deterministic fixed point. We now have $2L$ degrees of freedom, $x_i,y_i$ ($i=1,\dots,L$), and we will denote the corresponding fluctuations by $\boldvarphi=(\xi_1,\dots,\xi_L,\zeta_1,\dots,\zeta_L)$. 

Performing a Fourier transform in both position and time, one finds a linear equation of the form
\be\label{eq:2splang}
\mathbb{M}(k,\omega)\widetilde{\boldvarphi}(k,\omega)=\widetilde{\boldeta}(k,\omega).
\ee
The $2\times 2$ matrix $\mathbb{M}(k,\omega)$ and the properties of the white Gaussian noise $\boldeta$ can be obtained analytically (details are reported in the Appendix). The power spectra are then found from
\be\label{eq:p11p22}
P_{11}(k,\omega)=\avg{|\widetilde \xi(k,\omega)|^2}, ~~ P_{22}(k,\omega)=\avg{|\widetilde \zeta(k,\omega)|^2}.
\ee
We here remark that the presence of particles of type $2$ are irrelevant for the dynamics of the type-$1$ particles. The expression for $P_{11}(\omega)$ is therefore exactly the one one would obtain for a single-species TASEP on a ring.
\subsection{Test against simulations}
\begin{figure}[t]
\vspace{2em}
\centering
\vspace{-1em}
~~~~\includegraphics[scale=0.375]{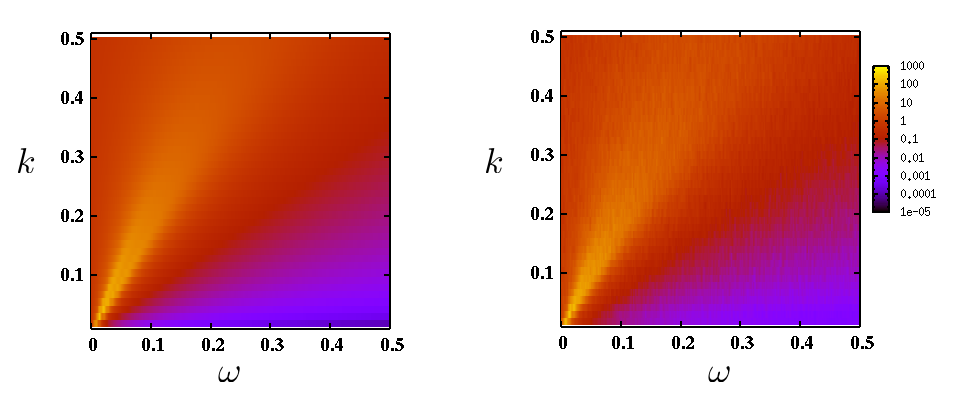} 
\vspace{0em}
\caption{(Colour online) Power spectrum $P_{22}(k,\omega)$ of the two-species TASEP. Parameters are as in Fig. \ref{fig:fig4}. The left-hand panel shows theoretical predictions for $P_{22}(k,\omega)$ (obtained from Eqs. (\ref{eq:p11p22})). The right-hand panel shows results from simulations at $\Omega=10$. Simulation data is averaged over $10$ independent runs. The colour indicates the magnitude of $P_{22}(k,\omega)$ in logarithmic scale (see legend).\label{fig:fig5}}
\end{figure}
We compare results of the theoretical computations against numerical simulations in Fig. \ref{fig:fig5}. As seen in the figure the qualitative agreement between theory and simulation is reasonable. We note that the power spectrum in the $k-\omega$ plane exhibits a `rim' on which most power is concentrated. In our sign conventions the rim is found at positive $k$ and $\omega$, and due to the symmetry $P(k,\omega)=P(-k,-\omega)$ at negative $k$ and $\omega$ as well. No significant concentration of power is to be expected when $k>0$ and $\omega<0$ or vice versa. This is at variance with the stochastic travelling waves observed in \cite{tommaso}, where peaks in the power spectrum are found in all four quadrants of the $k-\omega$-plane. It is here important to stress that the reaction-diffusion equations describing the system of \cite{tommaso} are of second order with respect to position, and hence they are invariant against reflections $x\to-x$. Waves travel in both directions in the model of \cite{tommaso}. In the exclusion processes this reflection symmetry no longer applies\footnote{The terms on the RHS of Eqs. (\ref{eq:2spmf}) can be written as first-order lattice derivatives of functions such as $x_i(1-x_{i+1})$ or $y_i(1-x_{i+1}-y_{i+1})$.}, particles generally travel to the right (the only exception in our model is the reaction $12\rightarrow 21$, when the particle of type $2$ effectively hops to the left), as also seen in Fig. \ref{fig:fig4}.

\begin{figure}[t!]
\vspace{2em}
\centering
\vspace{-1em}
~~~~\includegraphics[scale=0.55]{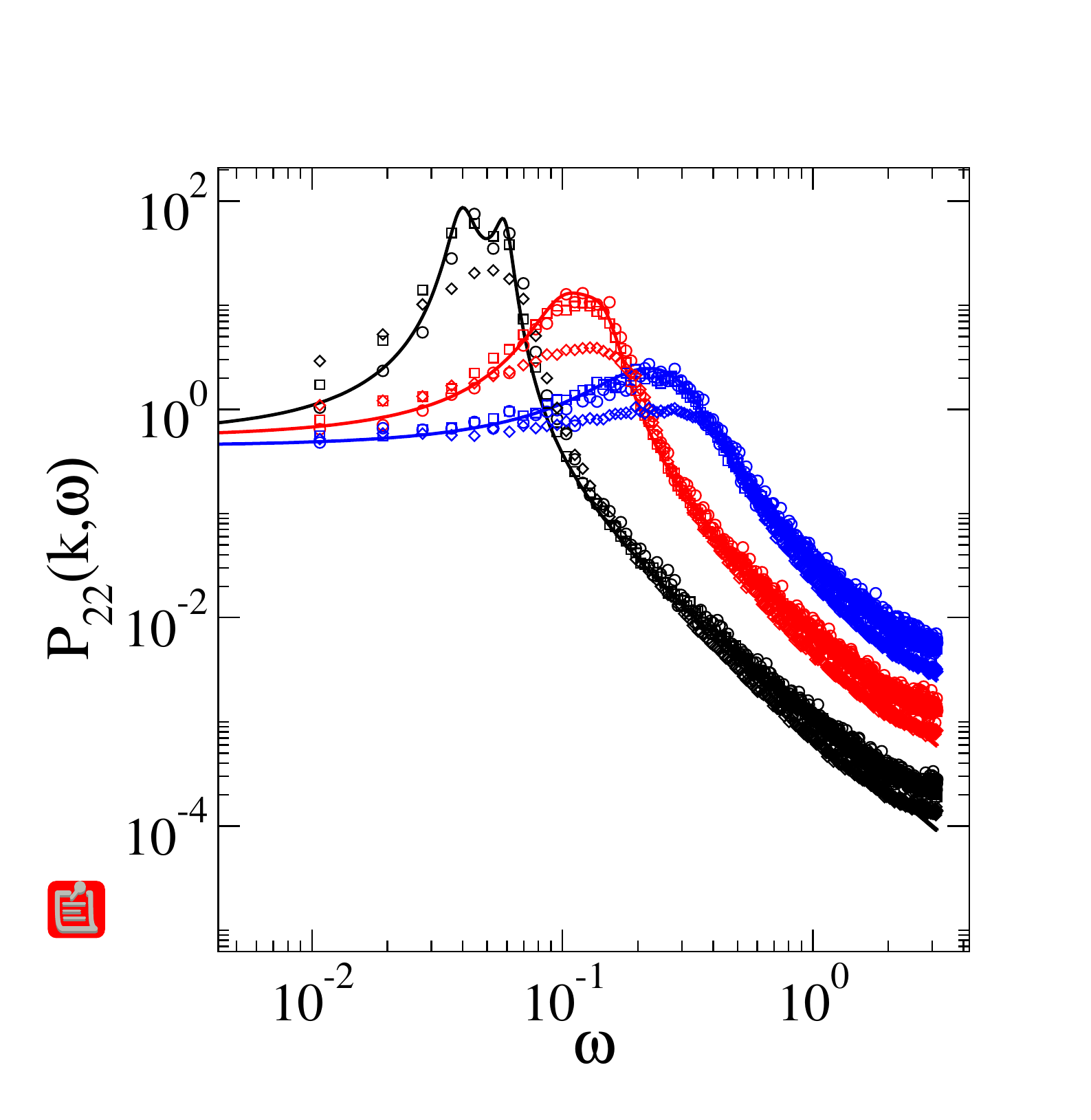} 
\vspace{0em}
\caption{(Colour online) Power spectrum $P_{22}(k,\omega)$ of the two-species TASEP at fixed values of $k=\frac{2\pi}{L}\ell$ with $\ell=2,3,4$ (left to right at the maximum of the curves). Parameters are $\rho_1=0.1,~ \rho_2=0.2$, the number of cells in the system is $L=128$. The solid lines show $P_{22}(k,\omega)$ as obtained from Eqs. (\ref{eq:p11p22}), markers are from simulations (circles: $\Omega=100$, squares: $\Omega=10$, diamonds: $\Omega=1$), averaged over $50-500$ independent runs, depending on the cell size. \label{fig:fig6}}
\end{figure}

In order to provide a more quantitative comparison between simulations and theory we depict the spectrum $P_{22}(k,\omega)$ as a function of $\omega$ at several fixed values of $k$ in Fig. \ref{fig:fig6}.

\section{Conclusions} \label{sec:concl}
In summary we have used a meta-population approach to describe several variants of the totally asymmetric exclusion process. The term `meta-population' here refers to a setup in which each cell of the underlying spatial structure can be occupied by several particles, up to a total capacity of $\Omega$. Carrying out a systematic expansion in powers of $\Omega^{-1/2}$ one derives the deterministic limiting equations in the leading order of the expansion, and obtains a set of Langevin equations describing fluctuations about this deterministic limit in the sub-leading order of the cell-size expansion. Such Langevin equations are not new for the description of exclusion processes, they have for example been formulated and used in \cite{zia0,zia2,cookdong}. These existing studies however take a mostly phenomenological approach, we feel that the angle taken here provides a more systematic derivation of an effective Langevin dynamics from first principles, and using well-controlled expansion techniques. Secondly, the meta-population model allows for a smooth interpolation between the mean-field limit and the standard single-occupancy ASEP model.  It is also important to stress that results from this expansion do not require any fitting parameters, unlike some of the more phenomenological approaches considered previously. While we report results mostly for $L=51$ cells in the system we note that our results are general and apply to any length of the ASEP chain. Our theory might therefore also be useful to shed more light on the $L$-dependence of fitting parameters and on the shape of power spectra reported e.g. in \cite{zia0,zia2}. We realize of course that the meta-population model is a-priori different from the conventional exclusion process, in which each cell can be occupied by at most one particle at any time.  However, as our results show the predictions derived from the sub-leading order of the expansion in the cell size agree reasonably well with simulations of the single-occupancy model\footnote{It should be noted that this agreement cannot generally be expected to hold very close to phase boundaries.}. We are therefore hopeful that the approach taken here might be useful to investigate other variants of the exclusion processes, for example two-lane models (see e.g. \cite{kolo1}) or models with spatial heterogeneities and/or individual `slow' sites \cite{cookdong}. Cell-size expansion techniques may also be considered for models with interactions reaching beyond neighbouring cells, such as for example the Nagel-Schreckenberg model of vehicular traffic \cite{nagel}. Work along these lines is in progress.

\section*{Acknowledgements} This work is partially funded by an RCUK Fellowship (RCUK reference EP/E500048/1), the author acknowledges support by EPSRC (IDEAS Factory - Game theory and adaptive networks for smart evacuations, EP/I005765/1). I would like to thank John Fry for useful comments on an earlier draft of the manuscript, and Tim Rogers for useful discussions. Helpful suggestions by two anonymous referees are gratefully acknowledged.

\appendix

\section{Spectra of the two-species process}
This appendix provides some more details of the calculation of power spectra in the two-species exclusion process discussed in Sec. \ref{sec:multi}. Similar to the standard TASEP model and the TASEP with constrained particle numbers one finds a set of Langevin equations of the type
\be
\dot \boldvarphi=\mathbb{J}\boldvarphi+\boldeta
\ee
to sub-leading order of the cell-size expansion. The matrix $\mathbb{J}$ is the $(2L)\times (2L)$ Jacobian of the deterministic dynamics, and $\boldeta$ describes a $2L$-component Gaussian noise variable. We will first calculate and simplify the Jacobian $\mathbb{J}$, and then address the noise.
\subsection{Calculation of the relevant Jacobian}
To simplify the notation we define $f_i$ and $g_i$ 
\BE
f_i\equiv -x_i(1-x_{i+1})+(1-x_i)x_{i-1}, \nonumber \\
g_i\equiv x_i y_{i+1}-x_{i-1}y_i+y_{i-1}(1-x_i-y_i)-y_i(1-x_{i+1}-y_{i+1})
\EE
as the expressions on the RHS of Eqs. (\ref{eq:2spmf}). Then one finds
\BE
\frac{\partial f_i}{\partial x_{i-1}}=1-\rho_1, ~~~~~ \frac{\partial f_i}{\partial x_{i}}=-1, ~~~~ \frac{\partial f_i}{\partial x_{i+1}}=\rho_1,\nonumber \\
\frac{\partial f_i}{\partial y_{i-1}}=0, ~~~~~ \frac{\partial f_i}{\partial y_{i}}=0, ~~~~ \frac{\partial f_i}{\partial y_{i+1}}=0,
\EE
 as well as
 \BE
\frac{\partial g_i}{\partial x_{i-1}}=-\rho_2, ~~~~~ \frac{\partial f_i}{\partial x_{i}}=0, ~~~~ \frac{\partial f_i}{\partial x_{i+1}}=\rho_2, \nonumber \\
\frac{\partial f_i}{\partial y_{i-1}}=1-\rho_1-\rho_2, ~~~~~ \frac{\partial f_i}{\partial y_{i}}=-1, ~~~~ \frac{\partial f_i}{\partial y_{i+1}}=\rho_1+\rho_2.
\EE
The $2\times 2$ Jacobian in Fourier space is therefore given by
\BE
\mathbb{J}(k)&=&\left(\begin{array}{cccc} (1-\rho_1)e^{-ik} -1 + \rho_1 e^{ik} & 0  \\ -\rho_2 e^{-ik}+\rho_2 e^{ik} & (1-\rho_1-\rho_2) e^{-ik}-1 + (\rho_1+\rho_2)e^{ik}   \end{array}\right)\nonumber \\
&=&\left(\begin{array}{cccc} \cos(k)-1+i(2\rho_1-1)\sin(k) & 0 \\ i2\rho_2\sin(k) & \cos(k)-1+i(2(\rho_1+\rho_2)-1)\sin(k)\end{array}\right).\nonumber \\
\EE
\subsection{Noise correlator}
Writing
\be
\avg{\eta_{a,i}(t)\eta_{b,j}(t')}=B_{ab,ij}\delta(t-t')
\ee
where $a,b\in\{1,2\}$ denotes the two species, and where $i,j\in\{1,\dots,L\}$ stands for cells we have
\BE
B_{11,ij}&=&(-T_2^*-T_4^*)\delta_{j,i-1}+(T_1^*+T_2^*+T_3^*+T_4^*)\delta_{ij}+(-T_1-T_3)\delta_{j,i+1}, \nonumber \\
B_{12,ij}&=&T_4^*\delta_{j,i-1}+(-T_3^*-T_4^*)\delta_{ij}+T_3\delta_{j,i+1}, \nonumber \\
B_{21,ij}&=&T_4^*\delta_{j,i-1}+(-T_3^*-T_4^*)\delta_{ij}+T_3\delta_{j,i+1}, \nonumber \\
B_{22,ij}&=&(-T_4^*-T_6^*)\delta_{j,i-1}+(T_3^*+T_4^*+T_5^*+T_6^*)\delta_{ij}+(-T_3-T_5)\delta_{j,i+1}, \nonumber\\
\EE
where we have used the shorthands
\BE
&&T_1^*=T_{i,1}^*, ~~~~~~~ T_2^*=T_{i-1,1}^*, \nonumber\\
&&T_3^*=T_{i,2}^*, ~~~~~~~T_4^*=T_{i-1,2}^*,\nonumber \\
&&T_5^*=T_{i,3}^*, ~~~~~~~T_6^*=T_{i,3}^*,
\EE
and where the asterisk indicates that these rates are to be evaluated at the deterministic fixed point.
Carrying out a Fourier transform with respect to position space (cell number) one finds
\BE
B_{11}(k)&=&(-T_2^*-T_4^*)e^{ik}+(T_1^*+T_2^*+T_3^*+T_4^*)+(-T_1-T_3)e^{-ik}, \nonumber \\
B_{12}(k)&=&T_4^*e^{ik}+(-T_3^*-T_4^*)\delta_{ij}+T_3e^{-ik},\nonumber \\
B_{21}(k)&=&T_4^*\delta_{j,i-1}+(-T_3^*-T_4^*)\delta_{ij}+T_3e^{-ik}, \nonumber \\
B_{22}(k)&=&(-T_4^*-T_6^*)e^{ik}+(T_3^*+T_4^*+T_5^*+T_6^*)+(-T_3-T_5)e^{-ik}. \nonumber\\
\EE
Inserting the fixed point values $T_1^*=T_2^*=\rho_1(1-\rho_1-\rho_2)$, $T_3^*=T_4^*=\rho_1\rho_2$ and $T_5^*=T_6^*=\rho_2(1-\rho_1-\rho_2)$ this becomes
\BE
B_{11}(k)&=&2\rho_1(\rho_1-1)\left[\cos(k)-1\right],\nonumber \\
B_{12}(k)&=&2\rho_1\rho_2\left[\cos(k)-1\right],\nonumber \\
B_{21}(k)&=&2\rho_1\rho_2\left[\cos(k)-1\right],\nonumber \\
B_{22}(k)&=&2\rho_2(\rho_2-1)\left[\cos(k)-1\right].
\EE

\subsection{Langevin equation and power spectrum}
The Langevin equation (\ref{eq:2splang}) is given by
\be
(i\omega-\mathbb{J}(k))\widetilde\boldvarphi(k,\omega)=\widetilde\boldeta(k,\omega),
\ee
where we have carried out Fourier transforms both with respect to position and time. This can be written as
\be\label{eq:lin}
\mathbb{M}(k,\omega)\widetilde\boldvarphi(k,\omega)=\widetilde\boldeta(k,\omega),
\ee
i.e.
\be
\mathbb{M}(k,\omega)=\left(\begin{array}{cc} i\omega-J_{11}(k) & -J_{12}(k) \\ -J_{21}(k) & i\omega-J_{22}(k)\end{array}\right),
\ee
i.e.
\BE
\hspace{-5em}\mathbb{M}=\left(\begin{array}{cccc} i\omega-\cos(k)+1-i(2\rho_1-1)\sin(k) & 0 \\ -i2\rho_2\sin(k) & i\omega-\cos(k)+1-i(2(\rho_1+\rho_2)-1)\sin(k)\end{array}\right).\nonumber \\
\EE
 Writing $\widetilde\boldvarphi=(\widetilde x,\widetilde y)$ and $\widetilde\boldeta=(\widetilde\eta_1,\widetilde\eta_2)$ the solution of Eq. (\ref{eq:lin}) is given by
\be
\left(\begin{array}{c} \widetilde{x}(\omega) \\ \widetilde{y}(\omega)\end{array}\right)=\frac{1}{\Delta(k,\omega)}\left(\begin{array}{cc} m_{22}(k,\omega) & -m_{12}(k,\omega) \\ -m_{21}(k,\omega) & m_{11}(k,\omega)\end{array}\right)\left(\begin{array}{cc} \widetilde \eta_1(\omega) \\ \widetilde\eta_2(\omega)\end{array}\right),
\ee
where
\be
\Delta=m_{11}m_{22}-m_{12}m_{21}.
\ee
From this we have the final result
\BE
\hspace{-4em}\avg{|\widetilde x(k,\omega)|^2}&=&\frac{1}{|\Delta|^2}\left[|m_{22}|^2b_{11}+|m_{12}|^2b_{22}-2\mbox{Re}[m_{22}m_{12}]b_{12}\right]\nonumber \\
\hspace{-4em}\avg{|\widetilde y(k,\omega)|^2}&=&\frac{1}{|\Delta|^2}\left[m_{11}|^2b_{22}+|m_{21}|^2b_{11}-2\mbox{Re}[m_{11}m_{21}]b_{21}\right],
\EE
where we have suppressed the dependencies on $k$ and $\omega$ on the RHS.

 \end{document}